\documentclass[showpacs,amsmath,twocolumn,superscriptaddress]{revtex4}
\usepackage{graphicx}
\usepackage{subfigure}
\usepackage[caption=false]{subfig}
\usepackage[all]{xy}
\usepackage{amsmath}
\usepackage{amssymb}
\usepackage{enumerate}
\usepackage{fancyhdr}

\def\and{{\rm and}}

\def\a{\alpha}
\def\b{\beta}

\def\d{\delta}

\begin{document}
\title{Holographic Superconductors in f(R) Gravity}

\author{Zhao Xue}
\email{xuezhao2012@pku.edu.cn}
\affiliation{Department of Physics and State Key Laboratory of Nuclear Physics and Technology,
Peking University, Beijing 100871, P. R. China}

\author{Sheng-liang Cui}
\email{shlcui@pku.edu.cn}
\affiliation{Department of Physics and State Key Laboratory of Nuclear Physics and Technology,
Peking University, Beijing 100871, P. R. China}

\begin{abstract}
We study the holographic superconductors in $f(R)$ gravity, and show how the critical temperature and condensate of the dual operators depend on the modifications of $f(R)$ to Einstein gravity. A nontrivial planar black hole which is asymptotic to AdS spacetime is investigated for a specific $f(R)$. The corrections to the thermal quantities of the black hole, the critical temperature and the condensate of the dual operator are all obtained in a perturbative approach explicitly. Some comments are given on the affections of such modifications to the holographic superconductors.

\end{abstract}


\maketitle

\section{introduction}
Motivated by the AdS/CFT correspondence~\cite{Maldacena9711200,Gubser9802109, Witten9802150}, the authors of~\cite{Horowitz0803.3295} constructed a holographic superconductor model, which presents a gravitational description of the (unconventional) superconductors. The Schwarzschild-AdS black hole provides a temperature to the system. A critical temperature $T_c$ exists, below which the black hole grows scalar hair, which corresponds to the condensate of the dual operator on the boundary. According to the AdS/CFT correspondence, the phenomena indicates there is a superconducting phase transition of the dual theory on the boundary~\cite{Horowitz0803.3295,Horowitz0810.1563}. Many works are carried out to study the topic, such as Gauss-Bonnet gravity~\cite{Gregory0907.3203,Pan0912.2475,Pan1005.4743,Barclay1009.1991,Gregory1012.1558,Barclay1012.3074,Cai1103.2833},
Born-Infeld theory~\cite{Jing1001.4227,Roychowdhury1201.6520,Roychowdhury1211.0904,Bai1212.2721}, or including both of them~\cite{Jing1012.0644,Roychowdhury1204.0673}. Despite the numerical results in the original paper~\cite{Horowitz0803.3295}, the phase transition can also be investigated analytically with the help of matching method~\cite{Gregory0907.3203} and Sturm-Liouville method~\cite{Siopsis1003.4275}.

The $f(R)$ gravity is one of modifications to the Einstein gravity, which has its advantage to generate the late-time expansion of the cosmic acceleration (see~\cite{Sotirious0805.1726} for a review). It is hard to find nontrivial black hole solutions with nonconstant Ricci scalar for general $f(R)$. There are even no such solutions in some cases~\cite{Whitt1984,Mignemi9202031}. Recently, several interesting solutions are obtained for some specific $f(R)$~\cite{Multamaki0606373,Olmo0612047}, even with matters~\cite{Sharif1302.1191}. Questions naturally arising here are whether we can find a suitable black hole to realize the superconducting phase transition, and how the modifications of $f(R)$ gravity affect the propertities of the corresponding holographic superconductors as already done in other higher order gravity theory
~\cite{Gregory0907.3203,Pan0912.2475,Pan1005.4743,Barclay1009.1991,Gregory1012.1558,Barclay1012.3074,Cai1103.2833,Jing1012.0644,Roychowdhury1204.0673}. According to the variousness of $f(R)$ theory, we only consider a specific $f(R)$ model to address the mentioned issues. Following the method~\cite{Multamaki0606373}, black holes of non-constant Ricci scalar with planar topology are obtained. Properties of the black hole and the holographic superconductors on the boundary are investigated in a perturbative manner. The corrections to the lowest order of $a$, which corresponds to the modifications of $f(R)$ to Einstein gravity as we will indicate, are obtained explicitly.

The paper is organized as follows. In section 2, we construct an asymptotic AdS black hole with planar topology and investigate the corresponding $f(R)$ theory. In section 3, we compute the modified critical temperature $T_c$ and the condensate of the dual operator $<\mathcal{O}^1>$. In the last section, we make a simple conclusion.
\section{A specific $f(R)$ model}
In order to realize a superconducting phase transition, nontrivial black holes which are asymptotic to AdS spacetime and have the planar topology are needed. In this section, we follow the method of~\cite{Multamaki0606373, Sharif1302.1191} to get such solutions and the corresponding $f(R)$. Action describing the $f(R)$ gravity and mini-coupled matter fields in four dimensional spacetime has the following form
\begin{eqnarray}\label{action}
S=\frac{1}{16\pi G}\int d^4x\sqrt{-g}\left [f(R)+\mathcal{L}_m \right],
\end{eqnarray}
where $f(R)$ is an analytic function of the Ricci scalar. When $f(R)=R-2\Lambda$, it recovers the Einstein gravity with cosmological constant. From the action, the vacuum modified Einstein equation can be written as
\begin{eqnarray}\label{Einstein}
R_{\mu\nu}f'(R)-\frac{1}{2}f(R)g_{\mu\nu}+(g_{\mu\nu}\nabla^2-\nabla_\mu\nabla_\nu)f'(R)=0,
\end{eqnarray}
where the prime on $f(R)$ means the derivative with respect to $R$. The equation usually includes higher derivatives of the metric for a general $f(R)$, which makes it difficult to solve. The general metric of the static and planar symmetric black holes have the following form:
\begin{eqnarray}
ds^2=-s(r)dt^2+p(r)dr^2+r^2(dx^2+dy^2).
\end{eqnarray}
As~\cite{Multamaki0606373, Sharif1302.1191}, solutions to the modified Einstein e.q.~(\ref{Einstein}) should satisfy the following equations:
\begin{eqnarray}\label{EOM6}
-2rF''+(r F'+2F)\frac{X'}{X}&=&0  , \nonumber\\
s''+(\frac{F'}{F}-\frac{X'}{X})(s'-\frac{2s}{r})-\frac{2s}{r^2}&=&0 ,
\end{eqnarray}
with definitions as $X(r)\equiv s(r)p(r)$ and $F(r)\equiv f'(R)$. We make an ansatz that $F(r)$ has the following form:
\begin{eqnarray}\label{F(r)}
F(r)=ar+b.
\end{eqnarray}
Substituting it into e.q.s~(\ref{EOM6}), solutions are obtained as follows:
\begin{eqnarray}
X&=&c_3 , \nonumber\\
s(r)&=&\frac{ac_2}{2b^2}-\frac{a^2c_2}{b^3}r-\frac{c_2}{3br} \nonumber\\
&&+r^2(c_1-\frac{a^3c_2}{b^4}log\frac{r}{ar+b}),
\end{eqnarray}
where $c_1,c_2,c_3$ are arbitrary constants. To have Schwarzschild-(A)dS black holes, we set $c_2=6bM$ and $c_3$ to be a positive number. Then the metric of the spacetime has the form
\begin{eqnarray}\label{metric0}
s(r)&=&\frac{3aM}{b}-\frac{6a^2M}{b^2}r-\frac{2M}{r} \nonumber\\
&&+r^2(c_1-\frac{6a^3M}{b^3}log\frac{r}{ar+b}),\nonumber\\
p(r)&=&\frac{c_3}{s(r)}.
\end{eqnarray}
The Ricci scalar can be calculated from the metric as follows:
\begin{eqnarray}\label{Ricci}
R(r)&=&\frac{-6abM}{c_3r^2(ar+b)^2}+\frac{24a^2M}{c_3r(ar+b)^2} \nonumber\\
&&+\frac{12(9a^3M-b^3c_1)}{bc_3(ar+b)^2}+\frac{24(3a^4M-ab^3c_1)r}{b^3c_3(ar+b)^2} \nonumber\\
&&-\frac{12a^2c_1r^2}{c_3(ar+b)^2}+\frac{72a^3M}{b^3c_3}log\frac{r}{ar+b}.
\end{eqnarray}
It's an analytic function of the radius $r$. In the asymptotic region with large $r$, the Ricci scalar can be expanded as a function of $\frac{1}{r}$
\begin{eqnarray}
R(r)|_{r\rightarrow +\infty}&=&-\frac{12c_1}{c_3}+\frac{72a^3M}{c_3b^3}log\frac{1}{a}+\mathcal{O}(\frac{1}{r^5}).
\end{eqnarray}
The result indicates that the spacetime is asymptotic to (A)dS. For latter discussion, we only consider the asymptotic AdS black holes, which can be obtained  by choosing the parameters properly. The effective cosmological constant and the radius of the asymptotic AdS spacetime can be defined as
\begin{eqnarray}\label{effective lambda1}
\Lambda_{eff}&=&-\frac{3c_1}{c_3}+\frac{18a^3M}{c_3b^3}log\frac{1}{a}, \nonumber\\
L_{eff}&=&\sqrt{\frac{-3}{\Lambda_{eff}}}.
\end{eqnarray}
Supposing the inverse function of e.q.~(\ref{Ricci}) can be obtained, we will get $F(r)$ as a function of $R$ by substituting $r(R)$ into the e.q.~(\ref{F(r)}). Then the corresponding $f(R)$ can be obtained from the integration of $F(R)$. Actually,
e.q.~(\ref{Ricci}) indicates that the exact inverse function can't be obtained. If $a=0$ in e.q.~({\ref{F(r)}}), then $F=b$ is a constant. The solutions of the e.q.s~(\ref{EOM6}) will be
\begin{eqnarray}\label{metric1}
s(r)&=&-\frac{2M}{r}+r^2c_1,\nonumber\\
p(r)&=&\frac{c_3}{s(r)},
\end{eqnarray}
which is just the trivial case with constant Ricci scalar. Then $f(R)$ can take any form as soon as it satisfies the constraint
$R f'(R)-2f(R)+3\square f'(R)=0$, which is obtained by contracting the e.q.~(\ref{Einstein}). Considering $a$ as a small parameter, the problem can be solved in a perturbative manner, and $a$ corresponds to the modifications of the $f(R)$ gravity to some extent. As we will see later, it relates to the parameter $\a$ which truly represents the modifications. The Ricci scalar (\ref{Ricci}) is expanded to the 1st order of $a$ as
\begin{eqnarray}
R(r)\thickapprox -\frac{12c_1}{c_3}-\frac{6aM}{c_3br^2}.
\end{eqnarray}
The inverse function is obtained as
\begin{eqnarray}\label{r(R)}
r(R)=\sqrt{\frac{-6aM}{12c_1b+c_3bR}}.
\end{eqnarray}
Substituting e.q.~(\ref{r(R)}) into e.q.~(\ref{F(r)}) and making a integration, we obtain the corresponding $f(R)$ theory to the 1st order of $a$,
\begin{eqnarray}\label{f(R)}
f(R)&=&\int dR F(R) \nonumber\\
&=&bR+c_4-2\sqrt{6}(\frac{a^3M}{c_3b})^{\frac{1}{2}}\sqrt{-12\frac{c_1}{c_3}-R}.
\end{eqnarray}
It is the approximate action, the higher order corrections of which are omitted.
The problem can also be viewed from the other point, which means that if we take~(\ref{f(R)}) as the precise action, the metric~(\ref{metric0}) can be checked to satisfy the equation of motion~(\ref{Einstein}) to the lowest order of $a$.  Now we make some justifications as follows:
\begin{itemize}
\item Compare e.q.~(\ref{f(R)}) with Einstein-Hilbert action with negative cosmological constant, we set $b=1$ and $c_4=-2\Lambda$;
\item From e.q.~(\ref{Ricci}), the limit is calculated as $\lim\limits_{a\rightarrow 0}R(r)=-\frac{12c_1}{c_3}$. So we set $\Lambda=-\frac{3c_1}{c_3}$ and $L^2=-\frac{3}{\Lambda}=\frac{c_3}{c_1}$;
\item The action shouldn't involve the integration parameters $M,c_1,c_3$ explicitly, so we define $\a=2\sqrt{6}(\frac{a^3M}{c_3b})^{\frac{1}{2}}$;
\item At last, we set $c_3=1$ for simplicity, which means we focus on a subclass of the solutions obtained above.
\end{itemize}
Under these justifications, the approximate $f(R)$ gravity has the following form,
\begin{eqnarray}\label{approximate action}
f(R)=R-2\Lambda-\alpha\sqrt{4\Lambda-R}.
\end{eqnarray}
The $f(R)$ theory of this type were already studied~\cite{Mazharimousavi2012} and nontrivial solutions with spherical topology were obtained ~\cite{Multamaki0606373}. When considering the AdS/CFT correspondence, the bulk gravity~(\ref{approximate action}) appears as classical dual of the quantum CFT on the boundary in large N limit. From the definition $\a\sim a^{\frac{3}{2}}$, we know that the parameter $a$ really represents the modification of $f(R)$
to Einstein gravity. The metric changes into
\begin{eqnarray}\label{metric2}
s(r)&=&3aM-6a^2Mr-\frac{2M}{r} \nonumber\\
&&+r^2(\frac{1}{L^2}-6a^3Mlog\frac{r}{ar+1}),\nonumber\\
p(r)&=&\frac{1}{s(r)}.
\end{eqnarray}
Note that in the Einstein limit $\alpha\sim 0$ or in the asymptotic region $r\sim \infty$, the solution truly goes to AdS spacetime.
The structure of the black hole is modified for nonzero $a$, which means the horizons are not degenerate anymore. For small $a$, the radius of the outer horizon can be calculated perturbatively. We assume the radius of the horizon has the following form:
\begin{eqnarray}
r_+=r_+^{(0)}+ar_+^{(1)}+a^2r_+^{(2)}+\mathcal{O}(a^3).
\end{eqnarray}
Substituting it into the equation $s(r_+)=0$, the corrected radius is obtained as follows to the 2rd order of $a$,
\begin{eqnarray}\label{horizon1}
r_+=r_+^{(0)}-\frac{a}{2}{r_+^{(0)}}^2-3a^2{r_+^{(0)}}^3,
\end{eqnarray}
with
\begin{eqnarray}
r_+^{(0)}=(2ML^2)^{\frac{1}{3}}.
\end{eqnarray}
As a quantum effect, the Hawking temperature of black holes is model independent.
It can be obtained either by the method of computing the surface gravity $\kappa$ or the Euclidean correspondence avoiding the conical singularity. For the metrc of this type, the temperature can be calculated simply as
\begin{eqnarray}\label{Hawking1}
T&=&\frac{s'(r)}{4\pi}|_{r=r_+} \nonumber\\
&=&\frac{3}{4\pi}(\frac{2M}{L^4})^{\frac{1}{3}}-\frac{9}{8\pi}a^2M+\mathcal{O}(a^3).
\end{eqnarray}
The result indicates that there are no corrections to the Hawking temperature to the 1st order of $a$. Other thermal quantities such as the entropy and energy, and the thermal dynamical laws should also be investigated. Black holes with planar topology in Einstein gravity were studied in~\cite{Cai1994,Cai9609065}. The form of boundary term and Euclidean method were addressed for $f(R)$ gravity~\cite{Dyer0809.4033,Guarnizo1002.0617}. But for the black holes with nonconstant curvature, it has some subtleties and still needs more investigations. So the topic is left to the further investigations. In latter discussions, $L=1$ is taken for simplicity.
\section{Condensate and critical temperature}
Following the method~\cite{Gregory0907.3203} and taking the probe limit as~\cite{Horowitz0803.3295,Horowitz0810.1563}, we discuss the holographic superconductors in the background described by the metric e.q.~(\ref{metric2}). The matter fields are chosen to be Maxwell field and charged complex scalar field, which are described by the following Lagrangian:
\begin{eqnarray}\label{Lagrangian}
\mathcal{L}_m=-\frac{1}{4}F_{\mu\nu}^2-|(\partial_\mu-i A_\mu)\psi|^2-m^2|\psi|^2.
\end{eqnarray}
The mass of the scalar field is taken to be $m^2=-\frac{2}{L_{eff}^2}$ with $L_{eff}$ defined in e.q.~(\ref{effective lambda1}). Referring to the symmetry of the background, we adopt the following ansatz:
\begin{eqnarray}
A_\mu=(\phi(r),0,0,0),~~~~\psi=\psi(r).
\end{eqnarray}
The equations of motion for matter fields are obtained as follows:
\begin{eqnarray}
\psi''+(\frac{s'}{s}+\frac{2}{r})\psi'+(\frac{\phi^2}{s^2}+\frac{2\gamma}{s})\psi&=&0 ,\label{EOM7}\\
\phi''+\frac{2}{r}\phi'-\frac{2\psi^2}{s}\phi&=&0 , \label{EOM8}
\end{eqnarray}
where the prime means the derivative with respect to $r$, and the parameter $\gamma$ is defined as $\gamma=1-6a^3Mlog\frac{1}{a}$.
Finiteness requires the electric potential to be zero $\phi(r_+)=0$ at the horizon. The e.q.s~(\ref{horizon1}) and (\ref{EOM7}) imply the relation $\psi'=-\frac{2}{3r_+^{(0)}}\psi$, which is calculated to the 1st order of $a$. In the asymptotic region, the solutions behave as
\begin{eqnarray}\label{BC1}
\phi(r)=\mu-\frac{\rho}{r},~~~~\psi(r)=\frac{\psi^{(1)}}{r}+\frac{\psi^{(2)}}{r^2},
\end{eqnarray}
where $\mu$ and $\rho$ are the chemical potential and the charge density of the dual field theory~\cite{Horowitz0803.3295,Horowitz0810.1563}. In 4-dimensions, both of the two terms of $\psi(r)$ are normalizable and can be viewed as the vacuum expectation values~\cite{Witten9905104}. Referring to the dictionary of holography, the vacuum expectation values of the dual scalar operators $\mathcal{O}^i$ on the boundary can be obtained as
\begin{eqnarray}\label{condensate}
<\mathcal{O}^i>=\sqrt{2}\psi^{(i)}.
\end{eqnarray}
Variable transformation is taken as $z=\frac{r_+}{r}$, under which the e.q.s~(\ref{EOM7}) and (\ref{EOM8}) change into
\begin{eqnarray}
\psi''(z)+\frac{s'}{s}\psi'(z)+\frac{r_+^2}{z^4}(\frac{\phi(z)^2}{s^2}+\frac{2\gamma}{s})\psi(z)&=&0 , \label{EOM9}\\
\phi''(z)-\frac{2r_+^2\psi(z)^2}{z^4s}\phi(z)&=&0 , \label{EOM10}
\end{eqnarray}
where all the primes mean the derivatives with respect to $z$. The asymptotic behavior e.q.~(\ref{BC1}) changes into
\begin{eqnarray}\label{BC4}
\phi(z)=\mu-\frac{\rho}{r_+}z,~~~~\psi(z)=\frac{\psi^{(1)}}{r_+}z+\frac{\psi^{(2)}}{r_+^2}z^2.
\end{eqnarray}
We expand the field $\psi(z)$ and $\phi(z)$ near the horizon $z\sim 1$,
\begin{eqnarray}
\psi(z)&=&\psi(1)-\psi'(1)(1-z)+\frac{1}{2}\psi''(1)(1-z)^2 \nonumber\\
&&+\cdots ,\label{Taylor}\\
\phi(z)&=&\phi(1)-\phi'(1)(1-z)+\frac{1}{2}\phi''(1)(1-z)^2 \nonumber\\
&&+\cdots. \label{Taylor1}
\end{eqnarray}
With the help of the equations of motion~(\ref{EOM9}) (\ref{EOM10}) and the relations at the horizon $\phi(1)=0$ and $\psi'(1)=\frac{2}{3}(1-\frac{a}{2}r_+^{(0)})\psi(1)$ to the 1st order of $a$, the Taylor expansions can be expressed with fewer parameters as
\begin{eqnarray}
\psi(z)&=&\psi(1)-\frac{2}{3}(1-\frac{a}{2}r_+^{(0)})\psi(1)(1-z) \nonumber\\
&&+(\frac{4}{9}-\frac{1}{4}ar_+^{(0)}
-\frac{\phi'(1)^2}{36{r_+^{(0)}}^2})\psi(1)(1-z)^2 \nonumber\\
&&+\cdots ,\label{Taylor4}\\
\phi(z)&=&-\phi'(1)(1-z)-\frac{1}{3}\psi(1)^2\phi'(1)(1-z)^2 \nonumber\\
&&+\cdots.\label{Taylor5}
\end{eqnarray}
In order to match the equations (\ref{BC4}) on the boundary and (\ref{Taylor4})~(\ref{Taylor5}) near the horizon, the following equations must be satisfied at some intermediate point $z_m$:
\begin{eqnarray}
\mu-\frac{\rho }{r_+}z_m&=&-\phi'(1)(1-z_m)\nonumber\\
&&-\frac{1}{3}\psi(1)^2\phi'(1)(1-z_m)^2 , \label{MC7}\\
-\frac{\rho}{r_+}&=&\phi'(1)+\frac{2}{3}\psi(1)^2\phi'(1)(1-z_m) , \label{MC8}\\
\frac{\psi^{(1)}}{r_+}z_m &=& \psi(1)-\frac{2}{3}(1-\frac{a}{2}r_+^{(0)})\psi(1)(1-z_m) \label{MC9}\\
&&+(\frac{4}{9}-\frac{1}{4}ar_+^{(0)}-\frac{\phi'(1)^2}{36{r_+^{(0)}}^2})\psi(1)(1-z)^2 ,\nonumber\\
\frac{\psi^{(1)}}{r_+} &=& \frac{2}{3}(1-\frac{a}{2}r_+^{(0)})\psi(1) \label{MC10}\\
&&-2(\frac{4}{9}-\frac{1}{4}ar_+^{(0)}-\frac{\phi'(1)^2}{36{r_+^{(0)}}^2})\psi(1)(1-z)^2. \nonumber
\end{eqnarray}
Here we only consider the case with nonvanishing $\psi^{(1)}$ to illustrate our purpose. Actually, a similar discussion can be carried out trivially for $\psi^{(2)}$. From e.q.s~(\ref{MC9}) and (\ref{MC10}), we obtain
\begin{eqnarray}
\frac{\phi'(1)}{r_+^{(0)}}&=&-6\sqrt{\frac{\frac{7}{9}-\frac{4}{9}z_m^2+(\frac{1}{12}+\frac{1}{4}z_m^2)ar_+^{(0)}}{(1-z_m^2)}}, \label{MC11}\\
\frac{\psi^{(1)}}{r_+}&=& \psi(1)[\frac{4+2z_m}{3(1-z_m)}-
\frac{2}{3}ar_+^{(0)}(1+z_m)]. \label{MC12}
\end{eqnarray}
When there is no condensate on the boundary, which means $\psi^{(1)}=0$, $\psi(1)$ must be zero.
From e.q.~(\ref{MC8}), we obtain
\begin{eqnarray}
\psi(1)^2=\frac{3}{2(1-z_m)}(-\frac{\rho}{r_+\phi'(1)}-1 ). \label{condensate4}
\end{eqnarray}
Using e.q.s~(\ref{horizon1}) (\ref{Hawking1}) (\ref{MC11}) and (\ref{condensate4}), we obtain the relation
\begin{eqnarray}\label{condensate5}
\psi(1)^2&=&\frac{3}{2(1-z_m)}(\frac{\b^2}{T^2}
+\frac{\b^2 \d a}{T}-1 ) \nonumber\\
&=&-\frac{3}{2(1-z_m)T^2}(T-\frac{\b^2\d a}{2}-\sqrt{\b^2+\frac{\b^4\d^2 a^2}{4}}) \nonumber\\
&&\times(T-\frac{\b^2\d a}{2}+\sqrt{\b^2+\frac{\b^4\d^2 a^2}{4}}),
\end{eqnarray}
where definitions are taken as $\beta^2=\frac{3}{2(4\pi)^2}\rho\sqrt{\frac{9(1-z_m^2)}{7-4z_m^2}}$ and $\delta=\frac{4\pi}{6}(\frac{11}{12}-\frac{1}{4}z_m^2)$.
When $T$ approaches some critical value $T_c$, the condensate $\psi^{(1)}$ approaches 0. So the critical temperature of the phase transition can be decided from e.q.s~(\ref{MC12}) and (\ref{condensate5}) as
\begin{eqnarray}\label{critical temperature2}
T_c&=&\frac{\b^2\d a}{2}+\sqrt{\b^2+\frac{\b^4\d^2 a^2}{4}} \nonumber\\
&=&\b+\frac{\b^2\d }{2}a+\mathcal{O}(a^2).
\end{eqnarray}
We define $T_c^{(0)}=\b=\lim\limits_{a\rightarrow 0}T_c$ as the critical temperature without $f(R)$ modifications. As discussed in~\cite{Horowitz0803.3295,Horowitz0810.1563,Gregory0907.3203}, the dual theory on the boundary turns into the superconducting phase below the critical temperature $T_c$. We plot the result when $a$ takes different values. From figure 1, we see that the modification shifts the curves of the critical temperature as a function of charge density $\rho$.
\begin{figure}
\centering
\includegraphics[width=2.5in]{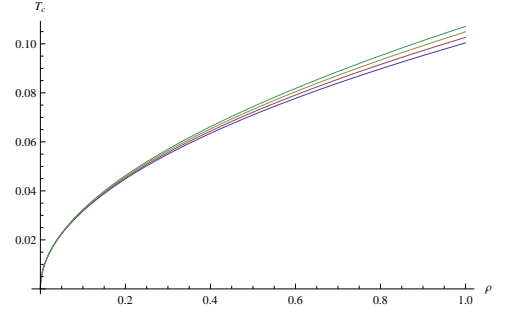}
\caption {$T_c$ varies with the parameter $\rho$, where the matching point is chosen to be $z_m=\frac{1}{2}$. The curves correspond to $a=0, \frac{1}{4}, \frac{1}{2}, \frac{3}{4}$ from the bottom up separatively.
}
\end{figure}
From e.q.s~(\ref{condensate}), (\ref{MC12}), (\ref{condensate5}) and (\ref{critical temperature2}), we obtain the condensate of the dual scalar operator on the boundary
\begin{eqnarray}\label{condensate6}
<\mathcal{O}^1>&=&\frac{4\pi}{3}\sqrt{\frac{3}{1-z_m}}\sqrt{(T_c-T)(T+T_c-\b^2\d a)} \nonumber\\
&&\times(\frac{4+2z_m}{3(1-z_m)}-\frac{2\pi(7-4z_m)}{9}aT )
\end{eqnarray}
Near the critical temperature $\frac{T}{T_c}\sim 1$, the condensate behaves as
\begin{eqnarray}\label{condensate7}
<\mathcal{O}^1>=\xi(1-\frac{T}{T_c})^{\frac{1}{2}},
\end{eqnarray}
with
\begin{eqnarray}
\xi&=&\frac{4\pi}{3}\sqrt{\frac{6}{1-z_m}}\left[\frac{4+2z_m}{3(1-z_m)}T_c-\frac{2+z_m}{6(1-z_m)}\b^2\d a  \right.\nonumber\\
&&\left.-\frac{2\pi(7-4z_m)}{9}T_c^2a \right].
\end{eqnarray}
We see that near the critical temperature it consists with the second order phase transition of superconductor, which is typical experimentally. Beside such region, the relation is modified by the parameter $a$ as e.q.~(\ref{condensate6}). When taking the limit $T\rightarrow 0$, the modified condensate at zero temperature is obtained as
\begin{eqnarray}\label{condensate8}
<\mathcal{O}^1>&=&\frac{4\pi}{3}\sqrt{\frac{3}{1-z_m}}\frac{4+2z_m}{3(1-z_m)}(T_c-\frac{1}{2}\b^2\d a)+\mathcal{O}(a^2) \nonumber\\
&=&\frac{4\pi}{3}\sqrt{\frac{3}{1-z_m}}\frac{4+2z_m}{3(1-z_m)}T_c^{(0)}+\mathcal{O}(a^2).
\end{eqnarray}
In the second line, the e.q.~(\ref{critical temperature2}) and the definition of $T_c^{(0)}$ are adopted.
The result shows that there are no corrections to the 1st order of $a$ when $T$ is zero. The expansion around zero temperature has the following form:
\begin{eqnarray}
<\mathcal{O}^1>&=&\frac{4\pi}{3}\sqrt{\frac{3}{1-z_m}}\frac{4+2z_m}{3(1-z_m)}T_c^{(0)}+\frac{4\pi}{3}\sqrt{\frac{3}{1-z_m}} \nonumber\\
&&\times\left[\frac{2+z_m}{3(1-z_m)}\b\delta-\frac{2\pi(7-4z_m)}{9}\b \right ]aT \nonumber\\
&&+\mathcal{O}(a^2).
\end{eqnarray}
From figure 2, we know that the presence of $f(R)$ makes it harder to form scalar hair either near zero temperature or the critical point.
\begin{figure}[h]
\centering
\subfigure[]
{\includegraphics[width=2.2in]{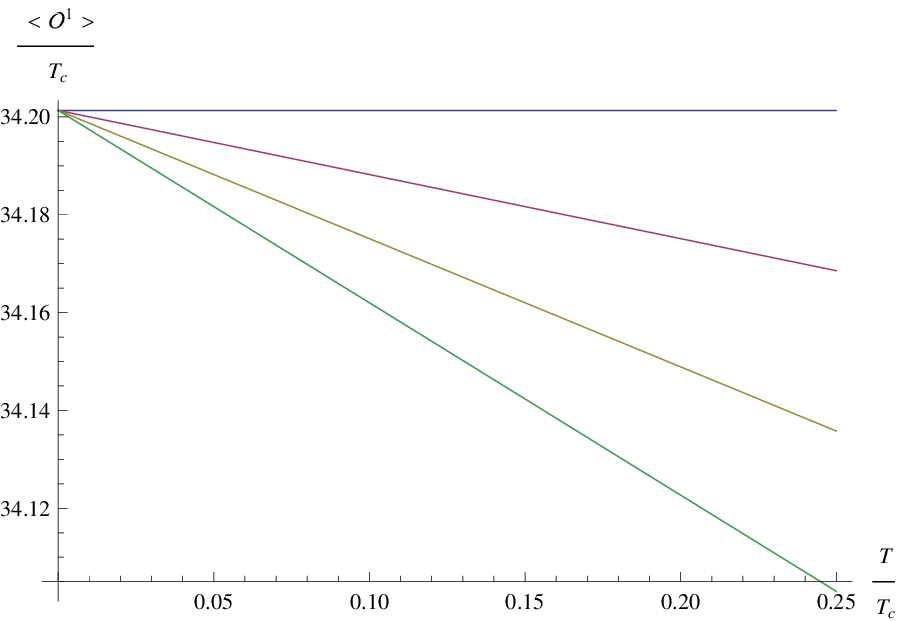}}
\subfigure[]
{\includegraphics[width=2.2in]{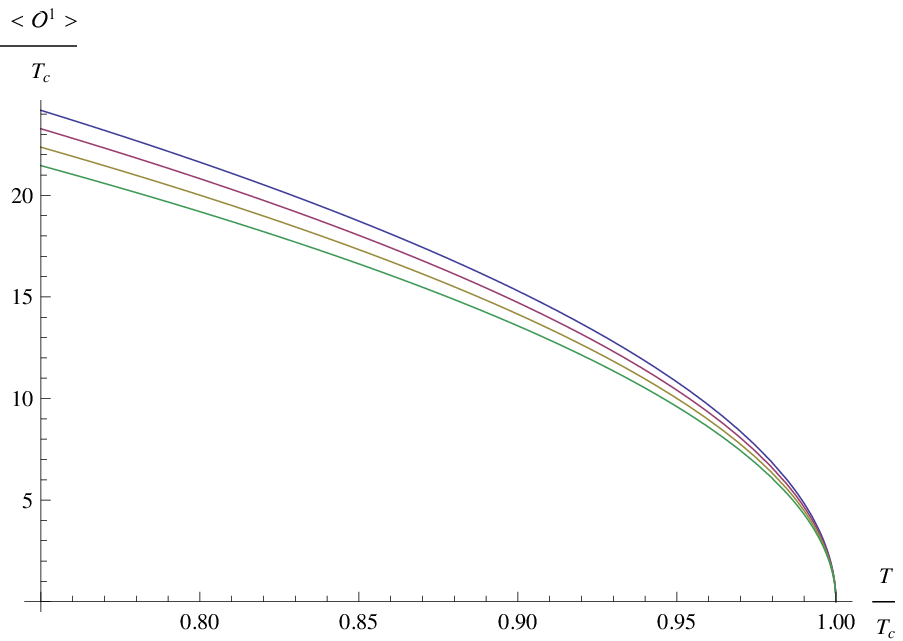}}
\caption {Figure (a) and (b) illustrate the condensates as a function of the temperature in different domains $0\leq \frac{T}{T_c}\leq\frac{1}{4}$
and $\frac{3}{4}\leq\frac{T}{T_c}\leq1$. In both pictures, the curves correspond to $a=0, \frac{1}{4}, \frac{1}{2}, \frac{3}{4}$ from up to down separatively. We choose the matching point $z_m=\frac{1}{2}$ and set $\rho=1$.
}
\end{figure}
We point it out again that the result is just referring to our specific model with approximate action~(\ref{approximate action}), and not universal to the $f(R)$ gravity. As a comparison, we can follow the same method to discuss the topic with the Schwarzschild-AdS black hole in Einstein gravity. The results are simply as follows:
\begin{eqnarray}
T'_c &=& \left(\frac{3}{2(4\pi)^2}\rho\sqrt{\frac{9(1-z_m^2)}{7-4z_m^2}} \right)^{\frac{1}{2}}, \\
<\mathcal{O}'^1>&=&\frac{4\pi}{9}\frac{4+2z_m}{1-z_m}\sqrt{\frac{3}{1-z_m}}\sqrt{T_c^2-T^2}.
\end{eqnarray}
Explicitly, they consist with our $f(R)$ model taking the limit $a\rightarrow 0$ in e.q.s~(\ref{critical temperature2}) and (\ref{condensate6}).
\section{Conclusions}
In this paper, we investigate the topic of holographic superconductors in $f(R)$ gravity. The background spacetime is an exact planar black hole solution, which is asymptotic to AdS spacetime for large radius. The corresponding $f(R)$~(\ref{approximate action}) is obtained in a perturbative manner for small $a$, which corresponds to the modifications referring to the relation $\alpha\sim a^{\frac{3}{2}}$. Thermal quantities and horizon structure are modified to the black hole, for which the perturbative results are obtained as~(\ref{horizon1}) and~(\ref{Hawking1}). From e.q.~(\ref{Hawking1}), we know the correction to the Hawking temperature is at least 2rd order. Following the method~\cite{Gregory0907.3203}, we compute the corrected critical temperature and condensate explicitly referring to the e.q.s~(\ref{critical temperature2}) and (\ref{condensate6}) separately. As illustrated in figure 1, we know the critical temperature of the superconductor is shifted. In figure 2, we see that it's more difficult to form the scalar hair for a given temperature in such an $f(R)$ theory. The calculations indicate that corrections are taking place even we investigate the topic to lowest order of $a$ only.

We would like to thank Bin Chen and Jia-ju Zhang for the helpful discussions. The work was in part supported by NSFC Grant No. 11275010.


\end{document}